# Elaboration d'entrepôts de données complexes

**Olivier Teste**

Université Paul Sabatier (Toulouse III),
IRIT (Institut de Recherche en Informatique de Toulouse), équipe SIG,
118, Route de Narbonne - 31062 Toulouse cedex 04, France
*tel* : (33) (0)5 61 55 63 22   *fax* : (33) (0)5 61 55 62 58
*mel* : Olivier.Teste@irit.fr
[Catégorie Jeune Chercheur]

**Résumé :**
Dans cet article, nous abordons le problème de la modélisation des entrepôts de données couramment utilisés dans les systèmes d'aide à la décision. Nous proposons un modèle permettant de décrire l'entrepôt comme un référentiel centralisé de données complexes, temporelles et extraites d'une source d'information. Notre modèle intègre trois concepts : l'objet entrepôt, la classe entrepôt et l'environnement. Chaque objet entrepôt est composé d'un état courant, de plusieurs états passés (modélisant les évolutions détaillées) et de plusieurs états archivés (modélisant les évolutions de manière résumée). Le concept d'environnement définit les parties temporelles dans le schéma de l'entrepôt avec une granularité pertinente (attribut, classe, graphe). Enfin, nous spécifions cinq fonctions visant à définir les structures de l'entrepôt et deux fonctions permettant d'organiser la hiérarchie d'héritage des classes entrepôt.

**Mots-clés :**
Modélisation des entrepôts, modèle orienté objet, données temporelles complexes.

**Abstract :**
In this paper, we study the data warehouse modelling used in decision support systems. We provide an object-oriented data warehouse model allowing data warehouse description as a central repository of relevant, complex and temporal data. Our model integrates three concepts such as warehouse object, environment and warehouse class. Each warehouse object is composed of one current state, several past states (modelling its detailed evolutions) and several archive states (modelling its evolutions within a summarised form). The environment concept defines temporal parts in the data warehouse schema with significant granularities (attribute, class, graph). Finally, we provide five functions aiming at defining the data warehouse structures and two functions allowing the warehouse class inheritance hierarchy organisation.

**Keywords :**
Warehouse modelling, object-oriented model, complex temporal data.

# 1 Introduction

L'approche des entrepôts de données ("*data warehousing*") est aujourd'hui unanimement reconnue comme étant une solution adaptée et performante, permettant d'améliorer la prise de décision dans les entreprises [WIDO95] [INMO96] [CHAU97] [GATZ99] [JARK99]. Un entrepôt se définit comme "*une collection de données intégrées, orientées sujet, non volatiles, historisées, résumées et disponibles pour l'interrogation et l'analyse*" [INMO96]. Il permet de stocker les données nécessaires à la prise de décision ; il est alimenté par des extractions de données portant sur des bases de production, appelées sources de données.

La présente étude poursuit les travaux effectués dans notre équipe en conception de système d'informations en milieu hospitalier [LAPU97] et vise à élaborer un système d'aide à la décision, basé sur l'approche des entrepôts de données, dans le milieu médical[i]. Plus précisément, notre système se propose d'améliorer l'analyse, le suivi et le contrôle des dépenses de santé, de l'activité des médecins et du "*comportement consommateur*" des patients. Nos travaux se placent dans le cadre du groupe EVOLUTION[ii], regroupant différentes équipes de recherche françaises, pour le développement de systèmes d'aide à la conception d'entrepôts.

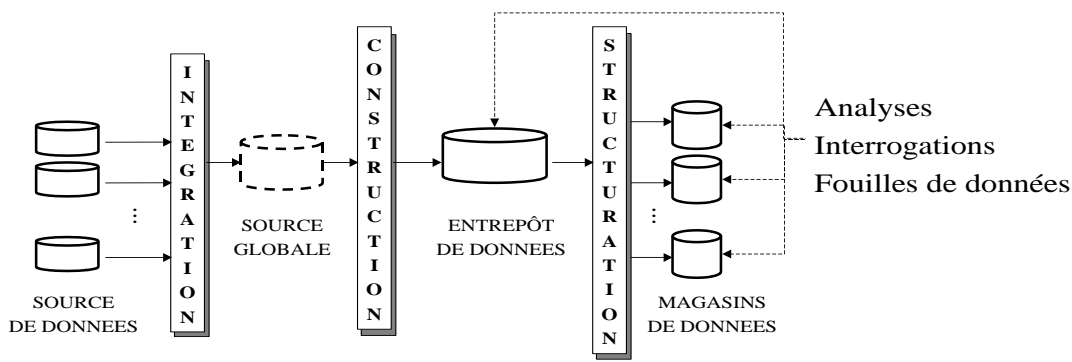

*figure 1 : Architecture du système d'aide à la décision.*

Nous avons défini un tel système [BRET99] [RAVA00], décomposé en trois niveaux (intégration, construction, structuration), dans le but de distinguer les différentes problématiques de recherche ; la figure 1 présente l'architecture du système.

- L'**intégration** se propose de résoudre les problèmes d'hétérogénéité (modèles, formats et sémantiques des données, systèmes,…) des différentes sources de données en intégrant celles-ci dans une source globale. Cette source globale est virtuelle, c'est à dire que les données utilisées pour la décision restent stockées dans les sources de données et sont extraites uniquement au moment des mises à jour de l'entrepôt.
Les techniques issues des bases de données fédérées [SAMO98] et réparties [RAVA96] peuvent être utilisées. De nombreux travaux de recherche traitent de ces problématiques, notamment le projet WHIPS[iii] [WIDO95] [LABI97], développé à l'université de *Stanford*, qui propose plusieurs algorithmes pour l'intégration et la maintenance des données issues de sources autonomes, distribuées et hétérogènes.

- La **construction** consiste à extraire les données pertinentes pour la prise de décision, puis à les recopier dans l'entrepôt de données, tout en conservant, le cas échéant, les changements d'états des données. L'entrepôt de données constitue une collection centralisée, de données

---

[i] Cette étude a été partiellement financée par le **CTI-Sud** (Centre de Traitement Informatique des régions Midi-Pyrénées et Languedoc-Roussillon) de l'Assurance Maladie.

[ii] Le projet **EVOLUTION** (http://www.prism.uvsq.fr/dataware/coop/evolution.html) se positionne dans le domaine de la conception des systèmes d'informations et propose le développement d'une méthodologie et d'un outil de type CASE pour l'aide à la conception et l'évolution des entrepôts de données.

[iii] http://www-db.stanford.edu/warehousing/index.html

matérialisées et historiques (conservation des évolutions), disponibles pour les applications de l'entrepôt.

Pour construire l'entrepôt, l'approche des vues matérialisées [GUPT95] est souvent utilisée. Les aspects techniques, comme la maintenance des vues matérialisées [HYUN97] [KOTI99] [YANG00] et la sélection des vues à matérialiser [THEO98] [THEO99] font l'objet de nombreuses propositions, tandis que les aspects modélisation restent peu abordés [GATZ99]. Récemment, des travaux abordent l'un des aspects essentiels des entrepôts : celui de la conservation des évolutions (données temporelles) ; [YANG00] propose un langage de définition de vues relationnelles temporelles et un algorithme de maintenance de ces vues matérialisées.

- La **structuration** réorganise les données dans des magasins de données afin de supporter efficacement les processus d'interrogation et d'analyse tels que les applications OLAP ("*On-Line Analytical Processing*" [CODD93]) et la fouille de données ("*Data Mining*" [FAYY96]). Pour ce faire, les données importées dans les magasins sont organisées de manière multidimensionnelle [AGRA97].

  De nombreux travaux proposent des modèles multidimensionnels, pour les bases relationnelles [AGRA97] [GYSS97] [LEHN98]. Ces différentes propositions sont parfaitement adaptées aux applications de gestion classiques, mais ne permettent pas de répondre complètement aux exigences des applications actuelles telles que les applications médicales [PEDE99]. En effet, ces dernières nécessitent des modèles plus riches que les modèles basés sur l'approche relationnelle afin de gérer des données complexes. Un autre intérêt de l'approche objet réside dans la modélisation des dimensions de l'analyse par des hiérarchies de composition pour supporter efficacement les analyses OLAP en niveaux de détails. [PEDE99] propose un modèle multidimensionnel orienté objet intégrant des données temporelles ainsi que des données imprécises. Cependant cette étude, située dans le milieu médical, se limite au dossier patient et ne propose pas de méthode pour constituer une base multidimensionnelle objet.

Dans notre architecture, au niveau de l'entrepôt on se concentre sur la gestion efficace des données extraites et sur la conservation de leurs évolutions, tandis qu'au niveau des magasins de données on se concentre sur les performances d'interrogation. Un magasin contient un sous ensemble des données de l'entrepôt traitant d'un métier particulier de l'entreprise [INMO96]. Les données relatives à un sujet à analyser sont donc réorganisées de manière adéquate, généralement de manière multidimensionnelle pour répondre aux exigences des applications OLAP. Dans [BRET99] nous avons proposé un modèle orienté objet multidimensionnel général, dédié aux magasins de données. Les données de l'entrepôt sont ainsi réorganisées au travers d'une classe de fait regroupant les mesures d'activité à analyser et de plusieurs classes de dimension correspondant aux différents paramètres de l'analyse. Cette proposition est validée par un outil d'aide à la réorganisation multidimensionnelle objet. D'autre part, nous avons mené une étude méthodologique pour concevoir des magasins de données orientées objet multidimensionnels [BRET00].

Cet article se positionne au second niveau de notre architecture et traitent de la construction de l'entrepôt de données. **Nous abordons la problématique de la modélisation et de l'élaboration de l'entrepôt.** A notre connaissance, il n'existe pas de proposition de modélisation des entrepôts basée sur le paradigme objet.

Notre étude aborde les points suivants :

- Nous proposons une modélisation de l'entrepôt permettant de décrire des données complexes et temporelles. Notre modèle prend en compte les besoins de modélisation de données complexes et intègre la dimension temporelle afin de conserver les évolutions des données de manière pertinente. Il doit être organisé de manière adaptée à une gestion efficace des données afin d'assurer la pérennité dans le temps de l'entrepôt.

- Nous présentons une solution visant à l'élaboration de l'entrepôt qui est construit par extractions de données issues de la source globale.

La section 2 étudie la modélisation de l'entrepôt et décrit notre modèle dédié aux entrepôt de données. La section 3 traite du processus d'élaboration de l'entrepôt, en proposant un mécanisme pour construire le schéma de l'entrepôt en utilisant des fonctions d'extraction, d'accroissement et de hiérarchisation. Enfin, la section 4 décrit brièvement un prototype visant à valider nos propositions.

## 2 Modélisation de l'entrepôt

L'entrepôt de données collecte des données, pertinentes pour supporter les processus de décision, issues d'une source globale. Ainsi, nous décrivons un exemple de schéma source dans le section 2.1. Ensuite, la section 2.2 présente le concept d'objet entrepôt et la section 2.3 celui de classe entrepôt. La section 2.4 introduit le concept d'environnement pour définir les parties temporelles de l'entrepôt. Enfin, la section 2.5 définit le schéma de l'entrepôt.

### 2.1 Cadre de l'étude : exemple de schéma source

La source globale, c'est à dire la base de données à partir de laquelle est élaboré l'entrepôt, est décrite avec un modèle de données orientées objet. Nous justifions notre choix par le fait que le paradigme objet s'est révélé parfaitement adapté pour l'intégration de sources hétérogènes [BUKH93] [RAVA96] [SAMO98] et permet de modéliser des sources intégrant des structures complexes.

Pour décrire les aspects statiques de cette source, nous utilisons le modèle objet standard : modèle de l'ODMG [CATT95] étendu par le concept de composition d'objets [BERT98].

Les travaux, présentés dans la suite de l'article, seront illustrés à partir de la source de données médicales suivante. Cette source de données décrit des patients et des praticiens. Ces derniers travaillent dans des services hospitaliers. Chaque service est dirigé par un praticien. Les patients consultent les praticiens, qui émettent des diagnostics et peuvent ordonner des analyses.

```
interface PERSONNE {
    attribute String nom;
    attribute String prénom;
    attribute Struct T_Adresse { String libelle,
                                 String ville,
                                 Long code_postal } adresse;
    attribute Short année_naissance;
    Short age();
    String département();
    }
interface PATIENT (extend PERSONNE) {
    attribute String no_insee;
    attribute String cle_insee;
    }
interface PRATICIEN (extend PERSONNE) {
    attribute String no_praticien;
    attribute String catégorie;
    attribute String spécialité;
    attribute Double revenus;
    relationship Set<SERVICE> travaille inverse SERVICE::équipe;
    relationship <SERVICE> dirige inverse SERVICE::est_dirigé;
    }
interface ETABLISSEMENT {
    attribute String nom;
    attribute String statut;
    attribute Struct T_Adresse { String libelle,
                                 String ville,
                                 Long code_postal } adresse;
    attribute Double budget;
```

```
      composition Set<SERVICE> organisation;
      }
interface SERVICE {
      attribute String nom;
      attribute String téléphone;
      relationship Set<PRATICIEN> équipe inverse PRATICIEN::travaille;
      relationship <PRATICIEN> est_dirigé inverse PRATICIEN::dirige;
      }
interface CONSULTATION {
      attribute Date date;
      attribute String commentaires;
      attribute String diagnostic;
      relationship Set<Image> analyses;
      relationship <PATIENT> patient;
      relationship <PRATICIEN> praticien;
      }
```

## 2.2 Concept d'objet entrepôt

Au niveau de l'entrepôt, chaque objet source extrait (ou groupe d'objets source) est représenté par un objet, appelé **objet entrepôt**. Un objet entrepôt est caractérisé par un identifiant interne, lequel est immuable, persistant au cours du cycle de vie de l'objet et indépendant des objets de la source.

L'entrepôt de données doit conserver les changements d'état des objets, tandis que la source de données ne contient généralement que l'état courant [CHAU97], ou bien, ne conserve qu'une partie récente des évolutions, insuffisante pour la prise de décision [YANG00]. Dans un entrepôt, l'administrateur peut décider de conserver :

- l'image de l'objet source, c'est à dire son **état courant**,
- les états successifs que prend l'objet source dans le temps, c'est à dire ses **états passés**,
- uniquement un résumé de ses états passés successifs, c'est à dire l'agrégation de certains états passés, appelée **états archivés**.

La figure 2 illustre notre proposition de modélisation des objets entrepôt, en représentant un objet entrepôt possédant un état courant, trois états passés et un état archivé.

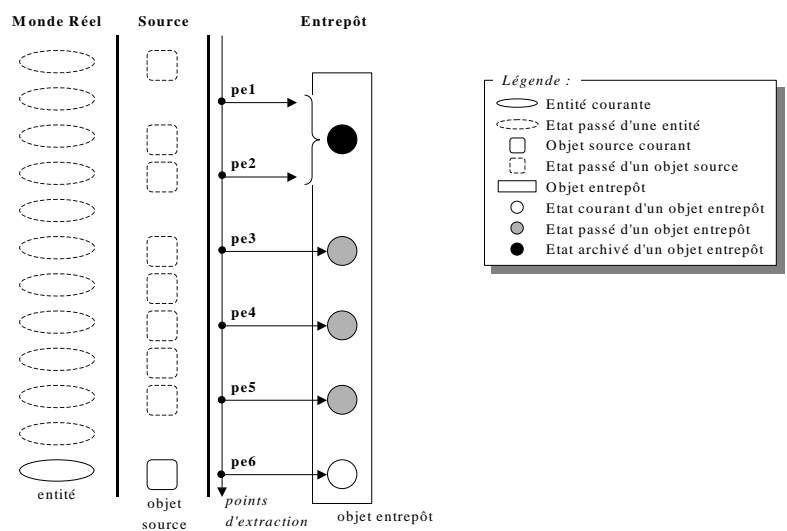

*figure 2 : Principe de modélisation d'un objet entrepôt.*

Deux approches sont envisageables pour rafraîchir l'entrepôt de données [WIDO95], c'est à dire pour répercuter dans l'entrepôt les évolutions qui surviennent au niveau de la source :

- L'approche du **rafraîchissement périodique** consiste à mettre à jour régulièrement l'entrepôt lors de points d'extraction correspondant à un état de cohérence de la source (terminaison des transactions).
- L'approche du **rafraîchissement incrémental** consiste à répercuter chaque modification d'une valeur de la source dès que possible. Cette approche fait l'objet de travaux importants dans le domaine des entrepôts [WIDO95] [ZHUG95] [YANG00].

Nous adoptons l'approche périodique puisque dans notre contexte (contrôle et suivi de l'activité des professionnels de santé et des patients) il est admis que l'analyse de l'activité des hôpitaux peut s'effectuer avec un décalage par rapport à la réalité. Ainsi l'état courant dans l'entrepôt ne correspond pas forcément à l'état courant de la source et certaines évolutions de la source peuvent ne pas être répercutées dans l'entrepôt (notamment lorsque plusieurs évolutions surviennent entre deux points d'extraction consécutifs). Ceci oblige l'administrateur à choisir une période de rafraîchissement de l'entrepôt adaptée aux besoins des applications décisionnelles et au rythme d'évolution de la source, tout en maintenant les performances de l'entrepôt lors de l'interrogation.

Nous formalisons le concept d'objet entrepôt par les définitions suivantes :

---
Un **objet entrepôt** $o$ est défini par le quadruplet ($oid$, $S_0$, $EP$, $EA$) où
- $oid$ est l'identifiant interne,
- $S_0$ est l'état courant,
- $EP = \{S_{p1}, S_{p2}, \ldots, S_{pn}\}$ est un ensemble fini contenant les états passés,
- $AP = \{S_{a1}, S_{a2}, \ldots, S_{am}\}$ est un ensemble fini contenant les états archivés.

---

Un objet entrepôt possède toujours un état courant, tandis que l'ensemble des états passés et l'ensemble des états archivés peuvent être vides.

---
Un **état** $S_i$ d'un objet entrepôt est défini par le couple ($h_i$, $v_i$) où
- $h_i$ est le domaine temporel correspondant aux instants durant lesquels l'état $S_i$ est courant,
- $v_i$ est la valeur de l'objet durant les instants de $h_i$.

---

Afin de définir le domaine temporel $h_i$ d'un état, notre modèle intègre un modèle temporel, linéaire, discret qui définit le temps par le biais d'unités de temps [GORA98]. L'espace continu du temps peut être représenté par une droite de réels, laquelle est décomposée en une suite d'intervalles consécutifs disjoints [WANG97] [FAUV99]. Chaque intervalle (régulier ou irrégulier) est non décomposable et tout instant de la droite de réels est approché par l'intervalle qui le contient. Les unités de temps sont caractérisées par la taille des intervalles décomposant la droite du temps. Notre modèle gère un ensemble d'unités temporelles nommées (*année*, *semestre*, *trimestre*, *mois*,…), muni d'une relation d'ordre partiel *est-plus-fine* permettant de comparer les unités. Enfin, nous définissons trois types temporels de base : durée, instant et intervalle.

---
Un **domaine temporel** $h_i$ est un ensemble ordonné d'intervalles, exprimé par $h_i = <[td^1, tf^1[; [td^2, tf^2[;\ldots;[td^h, tf^h[>$, satisfaisant aux propriétés suivantes :
- chaque intervalle est non vide, $\forall k \in [1..h]$, $td^k < tf^k$,
- les intervalles ont la même unité temporelle, $\forall k \in [1..h], \forall j \in [1..h]$, unit($[td^k, tf^k[$)=unit($[td^j, tf^j[$),
- les intervalles sont disjoints, $\forall k \in [1..h]$, $\forall j \neq k \in [1..h]$, $[td^k, tf^k[ \cap [td^j, tf^j[ = \varnothing$,
- les intervalles sont ordonnés et non contigus, $\forall k \in [1..h-1[, tf^k < td^{k+1}$.

---

La fonction $unit(Int)$ retourne le nom de l'unité temporelle utilisée par l'intervalle $Int$.

Le cycle de vie $[d, f]$ d'un objet entrepôt recouvre les domaines temporels de ses états passés, de ses états archivés et de son état courant :

$$<[d, f[> \supseteq h_0 \cup \left(\bigcup_{i=1}^{pn} h_i\right) \cup \left(\bigcup_{j=1}^{am} h_j\right).$$

Au cours de son cycle de vie, un objet entrepôt est initialement, **actif** tant que tous les objets sources à partir desquels il est généré, sont présents dans la source, puis, **gelé** lorsqu'au moins un de ses objets source a été supprimé. Ainsi, l'objet entrepôt peut être rafraîchi tant qu'il est actif, tandis qu'il n'est plus modifiable lorsqu'il est gelé, puisque son origine à la source n'est plus disponible.

**EXEMPLE :** La figure 3 représente graphiquement un objet entrepôt modélisant un hôpital. Cet objet a été créé le 1$^{er}$ janvier 1990 et il est rafraîchi chaque année. L'objet conserve les évolutions du budget et du nombre de services, sous une forme détaillée durant deux ans, puis sous une forme résumée pour les états les plus anciens. Il se compose d'un état courant, de deux états passés et d'un état archivé.

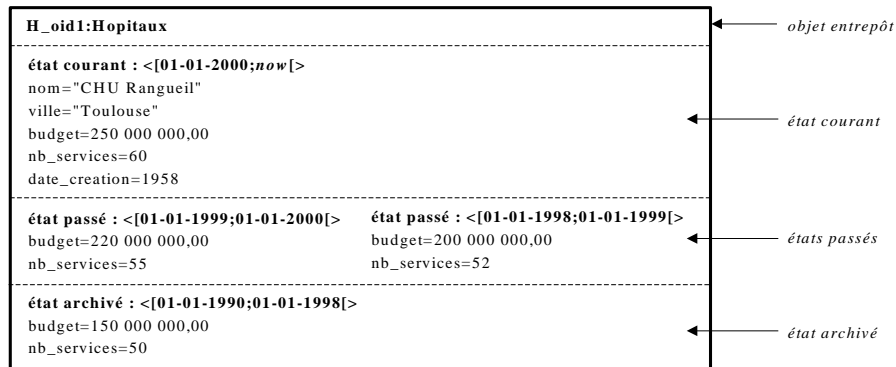

*figure 3 : Représentation graphique d'un objet entrepôt.*

## 2.3 Concept de classe entrepôt

Les objets entrepôt qui ont le même type abstrait, sont regroupés dans une classe. Nous définissons le concept de classe entrepôt comme suit.

Une **classe entrepôt** $c$ est définie par le septuplet ($Nom^c$, $Type^c$, $Super^c$, $Extension^c$, $Mapping^c$, $Tempo^c$, $Archi^c$) où
- $Nom^c$ est le nom de la classe,
- $Type^c$ est le type définissant la structure et le comportement des objets entrepôt de $c$,
- $Super^c$ est l'ensemble des super classes de c,
- $Extension^c = \{o_1, o_2,…, o_x\}$ est l'ensemble fini des objets entrepôt regroupés dans $c$,
- $Mapping^c$ est une **fonction de construction** qui permet de spécifier l'origine de la classe $c$,
- $Tempo^c$ est un **filtre temporel** définissant l'ensemble des propriétés temporelles de $c$ (une propriété est temporelle lorsque ses évolutions sont conservées par des états passés),
- $Archi^c$ est un **filtre d'archives** définissant l'ensemble des propriétés archivées de $c$ (une propriété est archivée lorsque ses évolutions passées sont résumées dans des états archivés).

Le type d'une classe $c$ est défini par un couple ($Structure^c$, $Comportement^c$) où $Structure^c$=($Attributs^c$, $Relations^c$) est un ensemble de propriétés, celles-ci pouvant être soit des attributs, soit des relations sémantiques (association, composition) entre la classe entrepôt $c$ et une (ou plusieurs) autre(s) classe(s), tandis que $Comportement^c$ est un ensemble d'opérations applicables.

Nous étendons le concept de type en spécifiant l'origine de chaque propriété.

- Une **propriété dérivée** est une propriété issue directement de la source. Le type et la valeur d'un attribut dérivé sont identiques à l'attribut source. Une relation dérivée représente une relation source ; elle est définie entre les classes entrepôt issues des classes source impliquées dans la relation source.
- Une **propriété calculée** est un attribut dont la valeur est le résultat d'une fonction (*avg*, *sum*, *max*, *min*, *count*,…) appliquée sur la source et dont le type est inféré par le résultat de cette fonction. Ce mécanisme permet à l'administrateur d'agréger des données source dans l'entrepôt.
- Une **propriété spécifique** est une propriété introduite par l'administrateur sans être issue de la source. Une telle propriété est valorisée soit par l'administrateur, soit par les utilisateurs qui peuvent ainsi compléter l'information contenue dans l'entrepôt.

De manière analogue, nous définissons une **opération dérivée** comme une opération issue de la source et une **opération spécifique** comme une opération ajoutée par l'administrateur (elle complète le comportement de l'entrepôt pour satisfaire des exigences liées aux applications décisionnelles).

Les classes entrepôt sont organisées suivant une hiérarchie d'héritage, modélisée par $Super^c$. Pour deux classes $c_i$ et $c_j$, $c_i$ est une sous classe de $c_j$ ($c_j$ est une super classe de $c_i$), notée $c_i \preccurlyeq c_j$ si et seulement si, $Type^{ci} \supseteq Type^{cj}$ et $Extension^{ci} \subseteq Extension^{cj}$.

La fonction de construction $Mapping^c$ permet de préciser comment est construite la classe entrepôt, c'est à dire, à partir de quelle(s) classe(s), avec quelle(s) propriété(s) dérivée(s), augmentée de quelle(s) propriété(s) spécifique(s) et de quel(s) attribut(s) calculé(s). Cette fonction, impliquée dans l'élaboration de l'entrepôt, fait l'objet d'une étude détaillée dans la section 3.

Les filtres temporel $Tempo^c$ et d'archives $Archi^c$, caractérisent respectivement les propriétés temporelles et d'archives de la classe. Par conséquent, le filtre temporel est constitué d'un ensemble d'attributs temporels dont les évolutions de valeur seront conservées par des états passés. Le filtre d'archives est un ensemble de couples (*attribut*, *fonction*) ; l'ensemble des attributs archivés (sous ensemble des attributs temporels de la classe) sont associés à une fonction d'agrégation qui indique comment sont résumées les évolutions détaillées dans les états d'archives.

**EXEMPLE :** Un administrateur désire élaborer un entrepôt contenant les informations relatives à la catégorie des praticiens chirurgiens. Il définit un entrepôt constitué de six classes, dont la définition est donnée ci-dessous, dans un langage de définition objet, proche de celui de l'ODMG.

- La classe entrepôt "*Personnes*" est constituée de propriétés dérivées et d'un filtre temporel précisant que les évolutions détaillées de l'attribut "*adresse*" sont conservées. Les évolutions des autres propriétés ne sont pas conservées.
- La classe "*Chirurgiens*" hérite de "*Personnes*". Son filtre temporel est composé de quatre propriétés tandis que son filtre d'archives contient uniquement les attributs temporels "*spécialité*" (résumé par la fonction *last*) et "*revenus*" (résumé avec la fonction *avg*).
- La classe "*Jeunes_Chirurgiens*" hérite de la classe "*Chirurgiens*".
- La classe "*Hôpitaux_Publics*" est constituée d'attributs dérivés ("*nom*", "*ville*" et "*budget*"), d'un attribut calculé ("*nb_services*"), d'un attribut spécifique ("*année_création*") et d'une relation de composition dérivée ("*organisation*"). Un filtre temporel et un filtre d'archives sont précisés.
- La classe "*Services*", composante de "*Hôpitaux_Publics*", comprend un attribut dérivé et deux relations d'association dérivées. Son filtre temporel indique les propriétés temporelles.
- La classe "*Etablissements*" hérite des classes "*Hôpitaux_Publics*" et "*Services*".

Nous précisons l'origine des propriétés par une lettre préfix "D_" pour dérivée, "C_" pour calculée et "S_" pour spécifique.

```
interface Personnes {
    D_attribute String nom;
    D_attribute String prénom;
    attribute Struct T_Adresse { String libelle,
                                  String ville,
                                  Long code_postal } adresse;
    D_attribute Short année_naissance;
    }
```

```
    with filters {
        temporal adresse;
        }
interface Chirurgiens (extend Personnes) {
        D_attribute Long no_praticien;
        D_attribute String spécialité;
        D_attribute Double revenus;
        D_relationship Set<Services> travaille inverse Services::équipe;
        D_relationship <Services> dirige inverse Services::est_dirigé;
        }
with filters {
        temporal spécialité, revenus, travaille, dirige;
        archive last(spécialité), avg(revenus);
        }
interface Jeunes_Chirurgiens (extend Chirurgiens) {
        }
interface Hôpitaux_Publics {
        D_attribute String nom;
        D_attribute String ville;
        D_attribute Double budget;
        C_attribute Short nb_services;
        S_attribute Short année_création;
        D_composition Set<Services> organisation;
        }
with filters {
        temporal budget, nb_services, organisation;
        archive avg(budget), avg(nb_services);
        }
interface Services {
        D_attribute String nom;
        D_relationship Set<Chirurgiens> équipe inverse Chirurgiens::travaille;
        D_relationship <Chirurgiens> est_dirigé inverse Chirurgiens::dirige;
        }
with filters {
        temporal équipe, est_dirigé;
        }
interface Etablissements (extend Hôpitaux_Publics, Services) {
        }
```

## 2.4 Environnement

Pour supporter efficacement les processus d'analyse décisionnelle, l'entrepôt de données doit être muni d'un mécanisme permettant de définir les parties temporelles, dont les évolutions de valeur seront conservées. En effet, les filtres associés aux classes entrepôt caractérisent comment sont résumés les évolutions de valeur des objets entrepôt, mais il est nécessaire de définir dans l'entrepôt les parties ayant un comportement temporel homogène (période de rafraîchissement, critères d'archivage,…) Pour cela, nous proposons le concept d'**environnement** pour définir ces parties temporelles cohérentes.

Un **environnement** *Env* est défini par le triplet ($Nom^{Env}$, $C^{Env}$, $Config^{Env}$) où
◆ $Nom^{Env}$ est le nom identifiant l'environnement,
◆ $C^{Env} = \{c_1, c_2,…, c_m\}$ est l'ensemble fini des classes contenues dans l'environnement,
◆ $Config^{Env}$ est un ensemble de règles de configuration, visant à définir différents paramètres locaux à l'environnement (période de rafraîchissement,…).

L'environnement regroupe des classes dont l'évolution de valeur des objets est conservée. Ainsi, les classes possédant un filtre sont contenues dans un environnement. Les environnements sont disjoints, $\forall Env^i, \forall Env^{j \neq i}, C^{Env_i} \cap C^{Env_j} = \emptyset$.

**EXEMPLE :** Reprenons l'exemple précédent. Les quatre classes "*Personnes*", "*Chirurgiens*", "*Hôpitaux_Publics*" et "*Services*" possèdent des filtres, ce qui oblige à la création d'au moins un environnement. L'administrateur définit un seul environnement regroupant ces quatre classes.

```
Environment Evolutions {
    class Personnes, Chirurgiens, Hôpitaux_Publics, Services;
}
```

Remarquer que la classe "*Chirurgiens*" hérite du filtre temporel de la classe "*Personnes*", car elles font partie du même environnement, tandis que la classe "*Jeunes_Chirurgiens*", qui n'appartient pas à l'environnement "*Evolutions*", hérite ni du filtre temporel de "*Personnes*", ni des filtres de "*Chirurgiens*".

L'héritage entre les classes d'un même environnement est étendu à leurs propriétés temporelles (filtres temporel et d'archives), $\forall c_i \in C^{Env}$, $\forall c_{j \neq i} \in C^{Env}$, $c_i \preccurlyeq c_j \Leftrightarrow Type^{c_i} \supseteq Type^{c_j} \wedge Extension^{c_i} \subseteq Extension^{c_j} \wedge \varphi^{c_i} \supseteq \varphi^{c_j} \wedge \psi^{c_i} \supseteq \psi^{c_j}$. Par contre, l'héritage entre deux classes n'appartenant pas au même environnement, ou n'appartenant pas à un environnement ne s'applique pas aux filtres, $\forall c_i \in C^{Env}$, $\forall c_j \notin C^{Env}$, $c_i \preccurlyeq c_j \Leftrightarrow Type^{c_i} \supseteq Type^{c_j} \wedge Extension^{c_i} \subseteq Extension^{c_j}$.

La configuration d'un environnement $Config^{Env}$ s'effectue au travers de règles ECA [DAYA88], spécifiées par l'administrateur, pour définir la gestion des objets entrepôt contenus dans l'environnement. Par exemple, l'administrateur définit des critères d'archivage (nombre d'états passés et/ou durée de conservation des états passés,…) ; nous avons étudié en détail les configurations dans [RAVA99].

Notre concept d'environnement autorise différents niveaux d'*historisation*. Contrairement à [YANG00] qui propose le seul niveau n-uplet pour l'*historisation* ou [PEDE99] qui propose d'*historiser* les données au seul niveau d'une table de dimension temps, nous proposons une *historisation* des données à trois niveaux : graphe, classe, attribut.

- Le **niveau classe** est le niveau intermédiaire qui consiste à restreindre un environnement à une seule classe entrepôt. Le filtre temporel de la classe sélectionne tous les attributs de la classe, garantissant ainsi la conservation de leurs évolutions. Notons que toutes les relations impliquant la classe ne sont pas *historisées*, à l'exception des relations réflexives.

- Le **niveau attribut** est le niveau le plus fin qui consiste à créer un environnement contenant uniquement une classe dérivée. Le filtre temporel précise les attributs temporels dont les évolutions doivent être conservées tandis que les attributs non sélectionnés par le filtre ne sont pas *historisés*.

- Le **niveau graphe** est le niveau le plus général où l'environnement est constitué d'un ensemble de classes. Ce niveau permet donc de conserver les évolutions des relations, en garantissant la conservation de leurs extrémités. Les relations reliant des classes de l'environnement sont *historisées*, tandis que les relations reliant une classe externe à l'environnement ne le sont pas. Rappelons que l'héritage entre les classes d'un même environnement est étendu aux propriétés temporelles des classes (filtres temporels et d'archives), tandis que l'héritage retenu pour ce qui concerne les classes externes à l'environnement, ne concerne que l'extension et le type.

## 2.5 Schéma de l'entrepôt

Un entrepôt se caractérise par son schéma défini comme suit.

---

Un **schéma d'entrepôt** $S^{ED}$ est défini par le quadruplet ($Nom^{ED}$, $C^{ED}$, $Env^{ED}$, $Config^{ED}$) où
- $Nom^{ED}$ est le nom de l'entrepôt,
- $C^{ED} = \{c_1, c_2,…, c_n\}$ est l'ensemble fini des classes de l'entrepôt,
- $Env^{ED} = \{env_1, env_2,…, env_p\}$ est l'ensemble des environnements de l'entrepôt,
- $Config^{ED}$ est un ensemble de règles de configuration, visant à définir les différents paramètres de l'entrepôt (période de rafraîchissement,…).

Les configurations de l'entrepôt $Config^{ED}$, de manière analogue à celles des l'environnements, déterminent le fonctionnement de l'entrepôt en terme de gestion des objets entrepôt. Ces règles globales sont appliquées à tout l'entrepôt, tandis que les règles locales à un environnement sont appliquées uniquement sur l'environnement. Dans un souci de simplification nous n'abordons pas ici cette problématique, que nous avons étudié dans [RAVA99].

Pour garantir la cohérence du schéma de l'entrepôt, nous introduisons une **contrainte d'intégrité**, permettant de vérifier que toutes les classes, extrémités des relations dérivées, sont présentes dans l'entrepôt. Ainsi, le schéma d'un entrepôt est **valide** si et seulement si toutes les relations dérivées de la source sont définies dans l'entrepôt, $\forall c_i \in C^{DW}$, $\forall r_k \in Relations^{c_i}$, $\exists c_j \in C^{DW} \mid r_k=(c_i, c_j, \tau)$ où $\tau='\alpha'$ pour une association et $\tau='\chi'$ pour une composition.

### 2.6 Synthèse

Dans cette section nous avons présenté notre modèle orienté objet pour les entrepôts de données complexes et temporelles. Nous avons défini les composants d'un schéma d'entrepôt pour répondre à notre problématique.

- Le **concept d'objet entrepôt** permet de stocker l'état courant d'une entité, ainsi que des états passés (évolutions détaillées) et des états archivés (évolutions résumées). Cette modélisation permet de conserver les données ainsi que leurs évolutions sous une forme pertinente avec un niveau de détail adéquat.
- Le **concept de classe entrepôt** étend le concept de classe en intégrant une fonction de construction, un filtre temporel et un filtre d'archives afin de prendre en compte les caractéristiques évolutives des objets entrepôt.
- Le **concept d'environnement** vise à définir les parties temporelles cohérentes dans l'entrepôt. Ce concept permet de définir trois niveaux d'*historisation* (graphe, classe et attribut) et étend l'héritage aux propriétés temporelles des classes.

La section suivante se propose d'étudier en détail la fonction de construction ($Mapping^c$) permettant l'élaboration des classes entrepôt.

## 3 Processus d'élaboration de l'entrepôt

L'entrepôt de données stocke des informations collectées à partir d'une source globale ; ce processus vise à construire le schéma de l'entrepôt, à partir de celui de la source globale. Nous proposons un processus de construction de l'entrepôt décomposé en deux phases :

- Une **phase d'extraction** qui consiste à définir la structure des classes entrepôt, en dérivant les propriétés des classes source pertinentes pour les utilisateurs, et éventuellement, en augmentant les classes entrepôt d'attributs calculés et de propriétés spécifiques.
- Une **phase de hiérarchisation** qui consiste à organiser la hiérarchie d'héritage des classes entrepôt, en fonction des besoins spécifiques de l'entrepôt liés aux applications décisionnelles.

### 3.1 Phase d'extraction

Nous proposons de caractériser la phase d'extraction au travers d'une fonction d'extraction, composée de fonctions de base, $f_1 \circ f_2 \circ \ldots \circ f_m$. On définit les fonctions de bases $f_i$ sur $cs$ dans $c$ ; par extension, $cs$ désigne l'ensemble des classes source, tandis que $c$ représente l'ensemble des classes entrepôt. Cependant, lorsque plusieurs fonctions $f_i$ composent l'extraction, on a $f_1 \circ f_2 \circ \ldots \circ f_m : c \rightarrow c_1 \rightarrow \ldots c_{m-1} \rightarrow c$ où $\forall i \in [1, m-1]$, $c_i$ est une classe temporaire.

Nous définissons les cinq fonctions de bases suivantes :

- La **fonction de projection** $\pi_P : cs \rightarrow c$ définit les propriétés dérivées (de l'état courant) de la classe entrepôt, à partir d'une classe source. L'ensemble des propriétés, spécifiées par $P=\{p_1, p_2, \ldots, p_m\}$, est un sous ensemble des propriétés de la classe source ($P \subseteq Structure^{cs}$).

- ◆ Extension$^c$={o $^{iv}$| ∃os∈Extension$^{cs}$, v$_0$=[p$_1$:os.p$_1$,…, p$_m$:os.p$_m$] avec ∀i∈[1,m], (p$_i$∈P ∧ p$_i$∈Structure$^{cs}$)},

    - ◆ Attributs$^c$={a | a∈P ∧ (a∈Attributs$^{cs}$ ∨ (a∈Attributs$^{cs'}$ | cs'∈Super$^{cs}$))},
    - ◆ Relations$^c$={r | r∈P ∧ (r∈Relations$^{cs}$ ∨ (r∈Relations$^{cs'}$ | cs'∈Super$^{cs}$))}.

- La **fonction de masquage** $\mu_P$ : $cs \rightarrow c$ (comparable à l'opérateur "*hide*" [ABIT91]), associée à l'ensemble de propriétés *P* (de la classe source) non dérivées, définit la structure de la classe entrepôt, composée des propriétés dérivées de la source n'appartenant pas à *P*.

    - ◆ Extension$^c$={o | ∃os∈Extension$^{cs}$, v$_0$=[p$_1$:os.p$_1$,…, p$_p$:os.p$_p$] avec ∀i∈[1,p], (p$_i$∉P ∧ p$_i$∈Structure$^{cs}$)},
    - ◆ Attributs$^c$={a | a∉P ∧ (a∈Attributs$^{cs}$ ∨ (a∈Attributs$^{cs'}$ | cs'∈Super$^{cs}$))},
    - ◆ Relations$^c$={r | r∉P ∧ (r∈Relations$^{cs}$ ∨ (r∈Relations$^{cs'}$ | cs'∈Super$^{cs}$))}.

- La **fonction d'accroissement** $\alpha_{P \times F}$ : $cs \rightarrow c$ consiste à créer de nouvelles propriétés dans la structure de la classe entrepôt générée. Ces propriétés P={p$_1$', p$_2$',…, p$_q$'} sont soit calculées ($f_i$∈{*count*, *sum*, *avg*, *max*, *min*} ou $f_i$∈Comportement$^{cs}$), soit spécifiques ($f_i$ est un type connu dans l'entrepôt).

    - ◆ Extension$^c$={o | ∃os∈Extension$^{cs}$, v$_0$=[p$_1$:os.p$_1$,…, p$_p$:os.p$_p$, p$_1$':$f_1$(os),…, p$_q$':$f_q$(os)]},
    - ◆ Attributs$^c$={a | a∈Attributs$^{cs}$ ∨ (a∈Attributs$^{cs'}$ | cs'∈Super$^{cs}$)} ∪ {p$_i$' | p$_i$'∈P, p$_i$' est un attribut},
    - ◆ Relations$^c$={r | r∈Relations$^{cs}$ ∨ (r∈Relations$^{cs'}$ | cs'∈Super$^{cs}$)} ∪ {p$_i$' | p$_i$'∈P, p$_i$' est une relation}.

- La **fonction de sélection** $\sigma_p$ : $cs \rightarrow c$ consiste à restreindre, par un prédicat de sélection *p*, l'extension de la classe source à partir de laquelle la classe entrepôt est générée.

    - ◆ Extension$^c$={o | ∃os∈Extension$^{cs}$, p(os) ∧ o.v$_0$ = os. v$_0$},
    - ◆ Attributs$^c$={a | a∈Attributs$^{cs}$ ∨ (a∈Attributs$^{cs'}$ | cs'∈Super$^{cs}$)},
    - ◆ Relations$^c$={r | r∈Relations$^{cs}$ ∨ (r∈Relations$^{cs'}$ | cs'∈Super$^{cs}$)}

- La **fonction de jointure** $\bowtie_p$ : $cs_1 \times cs_2 \rightarrow c$ consiste à filtrer le produit cartésien de deux classes source avec un prédicat de jointure *p*.

    - ◆ Extension$^c$ = {o | ∃os$_1$∈Extension$^{cs1}$, ∃os$_2$∈Extension$^{cs2}$, p(os$_1$,os$_2$) ∧ o.v$_0$ = [os$_1$.v$_0$, os$_2$.v$_0$]},
    - ◆ Attributs$^c$ = {a | (a∈Attributs$^{cs1}$) ∨ (a∈Attributs$^{cs'}$ | cs'∈Super$^{cs1}$) ∨ (a∈Attributs$^{cs2}$) ∨ (a∈Attributs$^{cs''}$ | cs''∈Super$^{cs2}$) },
    - ◆ Relations$^c$ = {r | (r∈Relations$^{cs1}$) ∨ (r∈Relations$^{cs'}$ | cs'∈Super$^{cs1}$) ∨ (r∈Relations$^{cs2}$) ∨ (r∈Relations$^{cs''}$ | cs''∈Super$^{cs2}$)}.

Notons que pour l'ensemble des fonctions de base, la classe entrepôt générée *c* n'a pas de super classe ($Super^c = \emptyset$).

**EXEMPLE :** Les classes entrepôt "*Chirurgiens*", "*Hôpitaux_Publics*" et "*Services*" définies dans la section 2.3 sont générées à partir des fonctions suivantes.

- La classe "*Chirurgiens*" est générée à partir d'une sélection des praticiens de la classe source "*PRATICIEN*" dont la catégorie est chirurgie. La structure générée contient toutes les propriétés de la classe source (les relations sont redéfinies entre les classes de l'entrepôt).

    **Mapping$^{Chirurgiens}$**=$\sigma_{p.catégorie="chirurgie"}$(p PRATICIEN)

- La classe "*Hôpitaux_Publics*" est construite à partir d'une sélection sur la classe source "*ETABLISSEMENT*" des hôpitaux publics. Une fonction de projection spécifie les propriétés dérivées ("*nom*", "*ville*", "*budget*", "*organisation*") et une fonction

---

[iv] Un objet entrepôt o=(*oid*, $S_0$, *EP*, *AP*) avec $S_0$=($h_0$, $v_0$) l'état courant, *EP* les états passés, *AP* les états archivés.

d'accroissement définit un attribut calculé ("*nb_services*") et un attribut spécifique ("*année_création*").

**Mapping$^{\text{Hôpitaux\_Publics}}$**=$\alpha$_{nb\_services:count(h.organisation), année\_création:Short}

$\pi$_{h.nom, h.adresse.ville, h.budget, h.organisation}

(h $\sigma$_{e.statut="public"}(e ETABLISSEMENT))

- La classe "*Services*" est construite en filtrant le produit cartésien entre les classes source "*ETABLISSEMENT*" (dont l'extension est restreinte aux hôpitaux publics) et "*SERVICE*". La fonction de masquage indique les propriétés non dérivées.

**Mapping$^{\text{Services}}$**=$\mu$_{sr.nom, sr.statut, sr.adresse, sr.budget, sr.téléphone}

(sr $\bowtie$_{h.organisation=s}

(h $\sigma$_{e.statut="public"}(e ETABLISSEMENT), s SERVICE)))

## 3.2 Phase de hiérarchisation

L'entrepôt de données doit être organisé en fonction des applications décisionnelles. La phase de hiérarchisation consiste à organiser la hiérarchie d'héritage dans entrepôt, en créant des super classes, et des sous classes entrepôt.

Lors de l'extraction, les classes entrepôt résultat ne sont pas organisées suivant une hiérarchie d'héritage (*Super$^c$*=$\varnothing$). Nous introduisons deux nouvelles fonctions visant à généraliser et à spécialiser les classes entrepôt.

- La **fonction de généralisation** $\Lambda_P$:$c_1 \times c_2 \times \ldots \times c_n \rightarrow c_0$ consiste à générer, à partir d'une ou plusieurs classes, une super classe regroupant l'ensemble $P=\{p_1, p_2, \ldots, p_m\}$ des propriétés communes aux classes, spécifiées par l'administrateur.

  ♦ Extension$^{c0}$=$\{o | \exists o' \in \bigcup_{j=1}^{n} Extension^{cj}, o.v_0=[p_1:o'.p_1,\ldots, p_m:o'.p_m], p_i \in P\}$,

  ♦ Structure$^{c0}$=$\{p | p \in \bigcap_{j=1}^{n} Structure^{cj} \wedge p \in P\}$,

  ♦ Super$^{c0}$=$\{c | c \in \bigcap_{j=1}^{n} Super^{cj}\}$,

  ♦ $\forall i \in [1..n]$, Structure$^{ci}$=$\{p | p \in Structure^{ci} \wedge p \notin P\}$,
  Super$^{ci}$=$\{c | c \in Super^{ci} \wedge c \notin \bigcap_{j \neq i} Super^{cj}\} \cup \{c_0\}$.

- La **fonction de spécialisation** $\Sigma_p$: $c_1 \times c_2 \times \ldots \times c_n \rightarrow c_0$ consiste à générer une sous classe à partir d'une ou plusieurs classes.

  ♦ Extension$^{c0}$=$\{o | \exists o_1 \in Extension^{o1},\ldots, \exists o_n \in Extension^{on}, p(o_1,\ldots o_n) \wedge o.v_0=[o_1.v_0,\ldots, o_n.v_0]\}$,

  ♦ Structure$^{c0}$=$\{p | p \in \bigcup_{j=1}^{n} Structure^{cj}\}$,

  ♦ Super$^{c0}$=$\{c_1, c_2,\ldots,c_n\}$.

**EXEMPLE :** Les classes entrepôt "*Personnes*", "*Jeunes_Chirurgiens*" et "*Etablissements*" définies dans la section 2.3 sont générées à partir des fonctions de spécialisation et de généralisation suivantes.

- La super classe "*Personnes*" regroupe les attributs "*nom*", "*prénom*", "*adresse*" et "*année_naissance*" de la sous classe "*Chirurgiens*".

  **Mapping$^{\text{Personnes}}$** = $\Lambda$_{c.nom, c.prénom, c.adresse, c.année\_naissance}(c Chirurgiens)

- La sous classe "*Jeunes_Chirurgiens*" spécialise la classe "*Chirurgiens*". Elle regroupe les jeunes chirurgiens dont la date de naissance est supérieure à 1970.

  **Mapping$^{\text{Jeunes\_Chirurgiens}}$** = $\Sigma$_{c.année\_naissance≥1970}(c Chirurgiens)

- La classe "*Etablissements*" spécialise les super classes "*Hôpitaux_Publics*" et "*Services*", en contenant les établissements hospitaliers situés dans la ville de Toulouse.

    **Mapping<sup>Etablissements</sup>** = $\Sigma_{e.organisation=s}$
    (e $\sigma_{h.adresse.ville="Toulouse"}$(h Hôpitaux_Publics), s Services)

# 4 Implantation

Dans l'optique de valider nos propositions, nous avons développé un prototype, appelé GEDOOH [v], acronyme de Générateur d'Entrepôts de Données Orientées Objet et Historisées [BRET99] [RAVA99]. Ce prototype se propose d'aider l'administrateur à élaborer le schéma d'un entrepôt de données. Il se compose de deux modules principaux.

- Une **interface graphique** visualisant le schéma de la source globale ainsi que celui de l'entrepôt,
- Un **générateur d'entrepôts** fournissant automatiquement les scripts de création de l'entrepôt, ainsi que les scripts de chargement initial et de rafraîchissement des données.

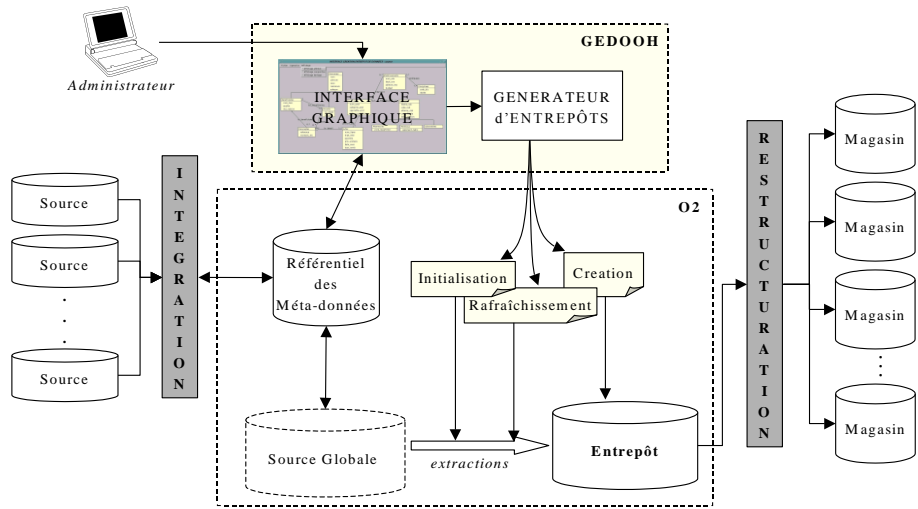

*figure 4 : Architecture du prototype GEDOOH.*

Notre outil facilite la tâche de l'administrateur pour construire et générer des entrepôts de données en proposant :

- Une représentation graphique des schémas de la source et de l'entrepôt.
- Une élaboration du schéma de l'entrepôt incrémentale et ergonomique grâce à un ensemble de fenêtres de saisie et de menus.
- Une simplification des éléments affichés, tels que les environnements (représentés par un double rectangle englobant les classes et les relations lui appartenant). Notons que les filtres temporels et d'archives sont explicitement représentées ; le symbole ' 🗔 ' définit une propriété appartenant au filtre temporel et le symbole ' 🗔 ' définit une propriété appartenant au filtre d'archives.
- Une génération automatique (à partir de la définition graphique du schéma de l'entrepôt) des scripts de création, d'alimentation et de rafraîchissement de l'entrepôt dans le système de gestion de base de données orientées objet (SGBDOO) hôte.

GEDOOH est un prototype opérationnel, implanté au dessus du SGBDOO O2. Il comporte approximativement 5000 lignes de code Java (jdk 1.2) pour l'interface et le générateur tandis que le référentiel des méta-données implanté dans O2, comprend environ 20 classes et représente 1500 lignes de code O2C.

---
[v] http://www.irit.fr/SSI/ACTIVITES/EQ_SIG/gedooh.html

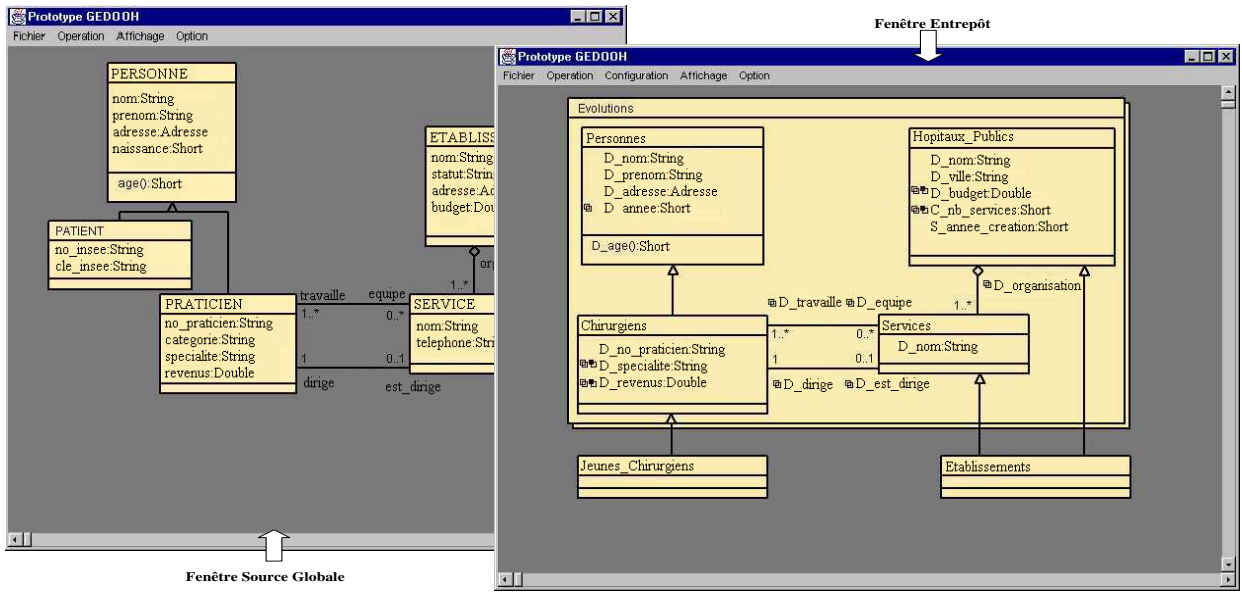

*figure 5 : Interface GEDOOH.*

## 5 Conclusion

Nous avons proposé une architecture pour les systèmes d'aide à la décision, fondée sur l'approche des entrepôts de données et distinguant différentes problématiques de recherche. Cet article étudie plus particulièrement l'élément central de notre architecture : l'entrepôt de données.

Premièrement, nous avons présenté un modèle d'entrepôt de données complexes et temporelles, basé sur le paradigme objet. Les principales contributions de notre modèle sont :

- Le concept d'objet entrepôt qui modélise l'état courant d'un objet source extrait, ainsi que des états passés (représentant les évolutions de l'objet sous une forme détaillée) et des états archivés (correspondant aux évolutions de l'objet décrites sous une forme résumée). L'intérêt de cette modélisation est de conserver les données de l'entrepôt ainsi que leurs évolutions à un niveau de détail pertinent.
- Le concept de classe entrepôt qui étend le concept de classe en intégrant les caractéristiques de notre approche par une fonction de construction, un filtre temporel et un filtre d'archives.
- Le concept d'environnement qui permet de définir simplement les parties temporelles homogènes dans le schéma de l'entrepôt. L'administrateur dispose de trois niveaux d'*historisation* (graphe, classe et attribut) pour spécifier les parties temporelles de l'entrepôt dans une taille adaptée aux besoins des applications de l'entrepôt.

Dans un second temps, nous avons proposé un processus d'élaboration de l'entrepôt, à partir d'un schéma de source globale. Ce processus repose sur

- des fonctions (projection, masquage, augmentation, sélection, jointure) visant à définir les structures de l'entrepôt et
- des fonctions (spécialisation, généralisation) permettant d'organiser la hiérarchie d'héritage des classes entrepôt.

L'ensemble de notre proposition est implantée au dessus du SGBDOO O2 dans le prototype GEDOOH [BRET99] [RAVA99]. Il permet de définir graphiquement le schéma de l'entrepôt à partir d'une représentation graphique de la source. Il génère les scripts de création des structures de l'entrepôt ainsi que les scripts de chargement initial et de rafraîchissement des données de l'entrepôt.

Notre processus d'élaboration de l'entrepôt n'intègre pas le comportement des classes. A l'heure actuelle, nous étudions le processus d'extraction automatique des méthodes. Ce processus nécessite d'analyser et de redéfinir la signature et le corps des méthodes dérivées dans l'entrepôt. Par ailleurs, il est indispensable de proposer un langage de manipulation des éléments de notre entrepôt. Ce langage doit englober une extension temporelle d'OQL [FAUV99] et prendre en compte les caractéristiques de nos objets entrepôt.

# 6 Remerciements

# 7 Références

[AGRA97]   Agrawal R., Gupta A., Sarawagi A., "Modeling Multidimensional Databases", ICDE'97.

[BERT98]   Bertino E., Ferrari E., Guerrini G., Merlo I., "*Extending the ODMG Object Model with Composite Objects*", OOPSLA'98, Vancouver (Canada), 1998.

[BRET99]   Bret F., Teste O., "Construction Graphique d'Entrepôts et de Magasins de Données", INFORSID'99, La Garde (France), Juin 1999.

[BRET00]   Bret F., Soule-Dupuy C., Zurfluh G., "Outil méthodologique pour la conception de bases de données décisionnelles orientées objet", LMO'00, St Hilaire (Canada), Jan. 2000.

[BUKH93]   Bukhres O.A., Elmagarmid A.K., "Object-Oriented Multidatabase Systems – A solution for Advanced Applications", Prentice Hall, ISBN 0-13-103813-2, 1993.

[CATT95]   Cattel R.G.G, "ODMG-93 Le Standard des bases de données objet", Thomson publishing, ISBN 2-84180-006-7, 1995.

[CHAU97]   Chaudhuri S., Dayal U., "An Overview of Data Warehousing and OLAP Technology", ACM SIGMOD Record, 26(1), 1997.

[CODD93]   Codd E.F., "Providing OLAP (on-line analytical processing) to user-analysts: an IT mandate", Technical Report EF Codd and Associate, 1993.

[DAYA88]   Dayal U., Blaustein B. T., Buchmann A. P., Chakravarthy U. S., Hsu M., Ledin R., McCarthy D. R., Rosenthal A., Sarin S. K., Carey M. J., Livny M., Jauhari R., "The HiPAC Project: Combining Active Databases and Timing Constraints", ACM SIGMOD Record, 17(3), Chicago (Illinois, USA), 1988.

[FAUV99]   Fauvet M.C., Dumas M., Scholl P-C., "A representation-independent temporal extension of ODMG's Object Query Language", BDA'99, Bordeaux, Oct. 1999.

[FAYY96]   Fayyad U.M., Piatetsky-Shapiro G., Smyth P., Uthurusamy R., "Advances in Knowledge Discovery and Data Mining", AAAI Press, ISBN 0-262-56097-6, 1996.

[GATZ99]   Gatziu S., Jeusfeld M.A., Staudt M., Vassiliou Y., "Design and Management of Data Warehouses", Report on the DMDW'99, ACM SIGMOD Record, 28(4), Dec. 1999.

[GORA98]   Goralwalla I.A., Özsu M.T., Szafron D., "An Object-Oriented Framework for Temporal Data Models", LNCS Temporal DBs, ISBN 3-540-64519-5.

| | |
|---|---|
| [GUPT95] | Gupta A., Mumick I.S., "Maintenance of Materialized Views: Problems, Techniques, and Applications", IEEE Data Engineering Bulletin, 1995. |
| [GYSS97] | Gyssen M., Lakshmanan L.V.S., "A Foundation for Multi-Dimensional Databases", VLDB'97, Athens (Greece), 1997. |
| [HYUN97] | Hyun N., "Multiple-View Self-Maintenance in Data Warehousing Environments", VLDB'97, Athens, 1997. |
| [INMO96] | Inmon W.H., "Building the Data Warehouse", John Wiley&Sons, ISBN 0471-14161-5. |
| [JARK99] | Jarke M., Lenzerini M., Vassiliou Y., Vassiliadis P., "Fundamentals of Data Warehouses", Ed. Springer Verlag, ISBN 3-540-65365-1, 1999. |
| [KOTI99] | Kotidis Y., Roussopoulos N., "DynaMat: A Dynamic View Management System for Data Warehouses", ACM SIGMOD'99. |
| [LABI97] | Labio W.J., Zhuge Y., Wiener J.L., Gupta H., Garcia-Molina H., Widom J., "The WHIPS Prototype for Data Warehouse Creation and Maintenance", SIGMOD, 1997. |
| [LAPU97] | Lapujade A., Ravat F., "Conception de systèmes d'information multimédia répartie : Application au milieu hospitalier", INFORSID'97, Toulouse, 1997. |
| [LEHN98] | Lehner W., Albrecht J., Wedekind H., "Normal forms for multidimensional databases", SSDBM'98, pp.63-72, 1998. |
| [PEDE99] | Pedersen T.B., Jensen C.S, "Multidimensional Data Modeling for Complex Data", ICDE'99, march 1999. |
| [RAVA00] | Franck Ravat, Olivier Teste, "Object-Oriented Decision Support System", To be appeared in the proceedings of ICEIS'00, July 4-7 2000, Stafford (UK). |
| [RAVA99] | Ravat F., Teste O., Zurfluh G., "Towards the Data Warehouse Design", ACM CIKM'99, Kansas City (Kansas, USA), Nov 1999. |
| [RAVA96] | Ravat F., "La fragmentation d'un schéma conceptuel orienté objet", Ingénierie des systèmes d'information, Vol 4, n°2, pp161-193, 1996. |
| [SAMO97] | Samos J., Saltor F., Sistrac J., Bardés A., "Database Architecture for Data Warehousing: An evolutionary Approach", DEXA'98, Vienna (Austria), 1998. |
| [THEO98] | Theodoratos D., Sellis T., "Data Warehouse Schema and Instance Design", ER'98, Singapore, 1998. |
| [THEO99] | Theodoratos D., Ligoudistianos S., Sellis T., "Designing the global data warehouse with SPJ views", CAISE'99, Heidelberg (Germany), June, 1999. |
| [WANG97] | Wang X.S., Bettini C., Brodsky A., Jajodia S., "Logical design for temporal databases with multiple granularities", ACM TODS, 22(2), 1997. |
| [WIDO95] | J. Widom, "Research problems in data warehousing", ACM CIKM'95, 1995. |
| [YANG00] | Yang J., Widom J., "Temporal View Self-Maintenance in a Warehousing Environment", EDBT'00, Konstanz (Germany), March 2000. |
| [ZHUG95] | Y. Zhuge, H. Garcia-Molina, J. Hammer, J. Widom, "View Maintenance in a Warehousing Environment", SIGMOD Record, San Jose (USA), 1995. |